\documentclass{IEEEcsmag}

\usepackage[colorlinks,urlcolor=blue,linkcolor=blue,citecolor=blue]{hyperref}
\usepackage{upmath,color}

\jvol{XX}
\jnum{XX}
\paper{8}
\jmonth{May/June}
\jname{BYU CS611 Project}

\pubyear{2019}

\begin{document}

\title{A brief history on Homomorphic learning: A privacy-focused approach to machine learning}

\author{Aadesh Neupane}




\begin{abstract}
Cryptography and data science research grew exponential with the internet boom. Legacy encryption techniques force users to make a trade-off between usability, convenience, and security. Encryption makes valuable data inaccessible, as it needs to be decrypted each time to perform any operation. Billions of dollars could be saved, and millions of people could benefit from cryptography methods that don't compromise between usability, convenience, and security. Homomorphic encryption is one such paradigm that allows running arbitrary operations on encrypted data. It enables us to run any sophisticated machine learning algorithm without access to the underlying raw data. Thus, homomorphic learning provides the ability to gain insights from sensitive data that has been neglected due to various governmental and organization privacy rules.  

In this paper, we trace back the ideas of homomorphic learning formally posed by Ronald L. Rivest and Len Alderman as ``Can we compute upon encrypted data?'' in their 1978 paper~\cite{rivest1978data}. Then we gradually follow the ideas sprouting in the brilliant minds of Shafi Goldwasser, Kristin Lauter, Dan Bonch, Tomas Sander, Donald Beaver, and Craig Gentry to address that vital question. It took more than 30 years of collective effort to finally find the answer ``yes'' to that important question.     


\end{abstract}

\maketitle
\begin{figure}
\centerline{\includegraphics[width=.9\linewidth]{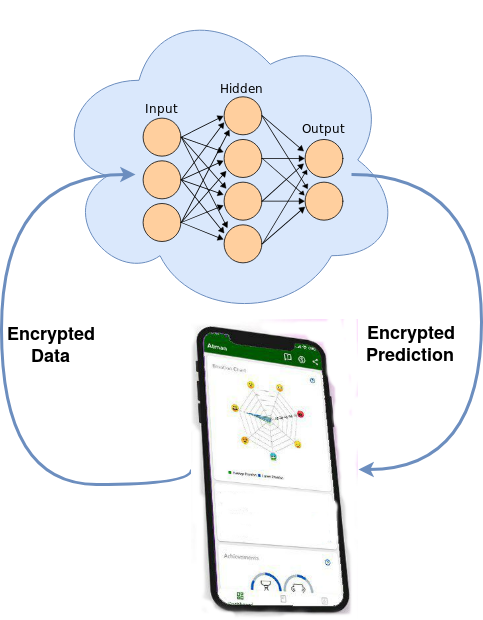}}
\caption{A simple system processing the user's sensitive information using homomorphic learning.}
\label{fig:simple_system}
\end{figure}

\chapterinitial{Homomorphic} learning is a cryptographic ``holy grail'' that allows a worker to perform arbitrary computations on client-encrypted data, without learning anything about the data itself. In simple terms, homomorphic learning is where a data owner wants to send data up to the cloud for processing but does not trust a service provider with their critical data. 
Using a homomorphic encryption scheme, the data owner encrypts their data and sends it to the server. The server performs the relevant computations on the data without ever decrypting it and sends the encrypted results to the data owner. The data owner is the only one able to decrypt the results since they alone have the secret key. Figure~\ref{fig:simple_system} shows a simple diagram that depicts the whole process involved in homomorphic learning. The paragraph below illustrates a few motivating examples of homomorphic learning.

Internet companies like Google, Facebook, and Amazon, along with healthcare companies like Accumen, AccuPoint, and Cigna collect huge amounts of data from users. Internet companies use sophisticated machine learning algorithms to learn patterns from the rich set of data and provide innovative user-focused services. Healthcare companies and government organizations are lagging behind others to mine the tons of data that could benefit millions of users and save billions of dollars due to strict privacy and data protection rules such as CCPA, GDPR, HIPPA, and FERPA. Homomorphic learning might be one way for these companies to mine sensitive data without violating privacy and data protection laws.       

Recently, the Dutch broadcaster VRT report~\cite{varhey2019} uncovered that the smart speakers (Amazon Echo, Google Home) are eavesdropping to every conversation even when not activated, and google employees are systematically listening to these recordings. Google spokesman said, ``This work is of crucial importance to develop technologies sustaining products such as Google Assistant.''  Homomorphic learning might be key for these companies in the future to improve AI technologies without access to the raw data.

There is an unprecedented amount of data breaches happening these days. But not every breach comes about as the result of outside attackers. There are other threat models to consider. Insider threats and privileged access account for about 35\% to 60\% of breaches, according to industry reports. 
Although the data collected by the companies is stored in the encrypted database using standard encryption (private key cryptography), the database administrators and developers still have access to all of the data. Tragically the raw data is released with the help of these private keys in the event of a data breach. Since homomorphic encryption allows learning without decrypting the data, there is no need for private keys. Thus, the use of homomorphic encryption could help to reduce data breaches.  

Homomorphic learning holds big promises for secure data storage and manipulation. It is far off from an efficient real-world enterprise implementation, but there has been substantial progress in the homomorphic learning area, such as secure multi-party computation and federation learning. It is both captivating and illuminating to witness the progress happening in this area. This paper tells the story of homomorphic learning: the web of people and ideas associated with it. Our story will begin from the time when Rivest asked the question ``Can we compute upon encrypted data?'' to ``fully homomorphic encryption for machine learning''~\cite{minelli2018fully}. Before describing the great question, we are going on a short detour to describe the history of modern cryptography.

\section{A Brief History of Cryptography}
There are historical records that the ancient civilization of the Incans, Aztecs, Greeks, and Romans practiced cryptography. Pieces of evidence suggest that their practice of cryptography was elementary. Though rudimentary cryptography was widely practiced until the seventeenth century, there were no formal tools to analyze the cryptography methods.

Johann Carl Friedrich Gauss (1777 - 1855), who had exceptional influence in many fields of mathematics and science, wrote an important statement about prime numbers which later became the foundation of modern cryptography. At the age of 21, in his famous book, ``Disquisitions Aritmeticae'' he wrote, ``The problem of distinguishing prime numbers from composite numbers and of resolving the latter into their prime factors is known to be one of the most important and useful in arithmetic.'' Most of modern cryptography is built on these two problems: finding if a given number is prime, and resolving composite numbers into their prime factors. 

The significant revolution in cryptography started with the World War, as the war introduced a qualitative change in patterns of communication. Radio communication was most popular during the war as the signals could be transmitted to very long distances. It could be used to give orders, convey strategies, and so on, but the enemy could easily intercept the line and discover what their strategies were. So this pitfall of radio provided a very strong demand for the use of cryptography. During World War II, special computing machines were used for cryptography. The Germans in particular had the famous enigma machine to encipher their messages. Alan Turing and others who had familiarity with notions of computation broke these ciphers, which had a significant impact on the war. The war was arguably shortened by several years because of these cryptographic breakthroughs.




Modern cryptography was only limited to military and government secret organizations, before the publication of David Kahn's book ``The Codebreakers - The Story of Secret Writing''~\cite{kahn1996codebreakers}. This book inspired general masses and lots of modern-day cryptographers, including Rivest, to become interested in the field. The NSA (National Security Agency), tried to suppress publication of the book because it brought attention to the field, and had a lot of interesting technical detail. But Kahn went ahead and published it anyway. He has updated the book since to include more recent developments.


Alan Turing was highly influential in the development of theoretical computer science, and one of his major contributions was to show that some class of problems were unsolvable on any computer; but then there were other classes of interesting problems which seem solvable in principle, but take a lot of time to solve. Thus he developed the theory of computational complexity. The problems in cryptography usually fall in the second class. So the hard problems are important for cryptography. Cryptographers want the problem to be very hard for the adversary to solve. 

Ralph Merkle, Marty Hellman, and Whit Diffie revolutionized cryptography by formalizing the idea of public-key cryptography in the paper ``New Directions on Cryptography''~\cite{diffie1976new}. Public-key cryptography was a marvelous idea which is qualitatively different notion than classical cryptography, because it separates the public key and the private key. 


The idea of public-key cryptography was on its way. Everybody would have a public key, and they could use that public key to encrypt a message. The public key would do the encryption, and a separate key (private) would do the decryption. Having these two keys be different supported by the notion of computational difficulty, made public-key cryptography secure. The core idea of public-key cryptography was that publishing one of the keys shouldn't reveal the other, and it should be computationally hard to figure out the secret key given the public key. So somebody could tell us the public key, and we wouldn't be able to figure out the secret key. 

The invention of the worldwide web, just like Marconi's invention of the radio, made cryptography essential and drove the demand for cryptography.\footnote{All the content of this section is based on the talk by Ronald L.Rivest at the award ceremony for Marconi Prize in 2007}

\section{RSA's Dream}
\textbf{RSA} (Rivest–Shamir–Adleman) is one of the most popular public-key cryptosystems and is widely used for secure data transmission. Ronal Linn Rivest is one of the authors of the \textbf{RSA}~\cite{rivest1978method} paper that introduced a public-key cryptosystem. He grew up in Niskayuna, New York, a suburb of Schenectady, New York where he attended public school. He graduated from Niskayuna High School in 1965. He received a B.A. in Mathematics from Yale University in 1969. The David Kahn's book ``The Codebreakers - The Story of Secret Writing''~\cite{kahn1996codebreakers}  made him interested in cryptography during his undergraduate days. 

During his Ph.D., he briefly worked in the area of Artificial Intelligence, but soon realized that his real passion was for mathematics and theoretical computer science. He received a Ph.D. in Computer Science from Stanford University in 1974. His Ph.D. research was mainly on the analysis of associative retrieval algorithms~\cite{rivest1974analysis} under supervisor Professor Robert Floyd. After appointment as a faculty in the Department of Computer Science at MIT, we are fortunate that he got a chance to communicate with Ralph Merkle, Marty Hellman, and Whit Diffie about public-key cryptography. Their paper ``New Directions on Cryptography''~\cite{diffie1976new} had a long-lasting impact on Rivest that inspired him to work with Shamir and Adleman to formulate \textbf{RSA} in their seminal paper ``A method for obtaining digital signatures and public-key cryptosystems''~\cite{rivest1978method}.



\begin{figure}
\centerline{\includegraphics[width=18.5pc]{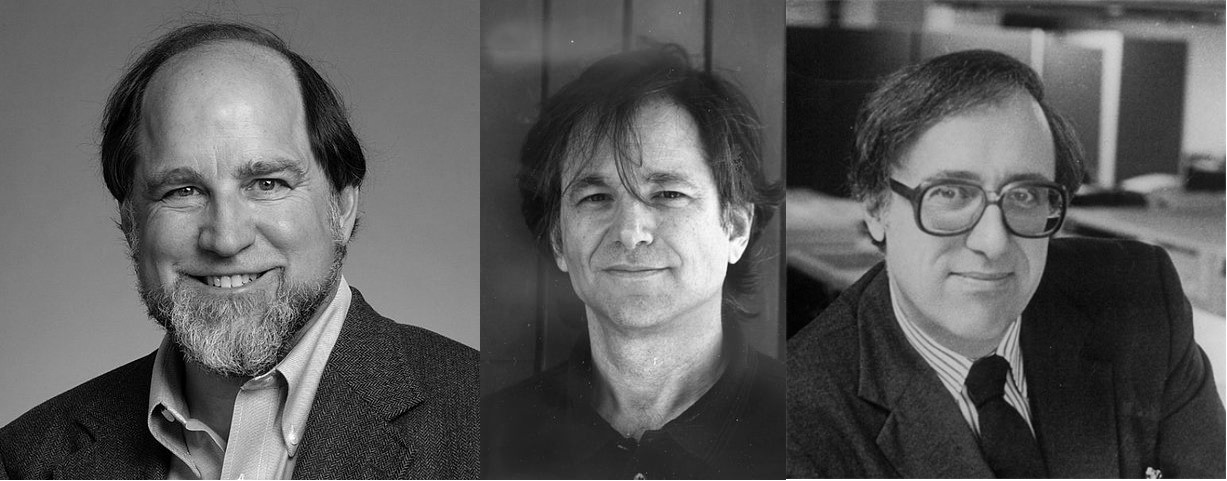}}
\caption{Computer scientist who first thought about privacy homomorphism. From left to right: Ronal Linn Rivest, Len Adleman, and Mike Dertouzos}
\label{fig:Scientist}
\end{figure}

With the success of public-key cryptosystems, Rivest was dreaming about doing computations with  encrypted data. So just after the publication of their seminal paper, Len Adleman and Ronal Linn had a fruitful discussion with Mike Dertouzos, who was the Director of the MIT Laboratory for Computer Science (LCS). During the discussion they asked him the, ``\textbf{Can we compute upon encrypted data?} Can we work with data that's been encrypted piece by piece and combine the pieces while leaving them encrypted so we can end up taking two ciphertexts, combine them, and end up with a new ciphertext in such a way that the underlying plain text for that new ciphertext is the appropriate operation on the underlying plain text?'' 

They found out that one way of answering their question was with a process called \textit{homomorphism}, and they dreamed of a cryptosystem that's \textit{homomorphic}. They imagined a cryptosystem with a sufficiently rich set of operations, that supports arbitrary computations on encrypted data. But they had no idea on how to pursue this dream. They get credit for asking this important question: but they didn't solve it. They formally asked this question in their paper ``On data banks and privacy homomorphisms''~\cite{rivest1978data}. They defined the term ``privacy homomorphism'' as a special encryption function which permits encrypted data to be operated on without preliminary decryption of the operands, for many sets of interesting operations.  

The ``privacy homomorphism'' idea developed in their famous paper was borrowed from mathematics. \textit{Homomorphic} is an adjective that describes something of the same or similar form. Specifically, in algebra, a \textit{homomorphism} is a structure-preserving map between two algebraic structures of the same type (such as two groups, two rings, or two vector spaces).

\section{Spring Break}  
The question Rivest et al.~\cite{rivest1978data} asked in 1978 about computation on encrypted data was way ahead of its time. It was too abstruse to be answered as a whole with the current state of research in cryptography. There was a long silence in the research community. It felt like the researchers were on a long spring vacation. Every four/five years, sporadically, a few researchers would pursue somewhat related topics to ``privacy homomorphism''. It took more than thirty years for the community to catch up with the question.

During this ``spring break'', Shafi Goldwasser was one of the prominent researchers who formalized vital ideas like probabilistic encryption and zero-knowledge interactive proofs. Her paper on \textit{probabilistic encryption} was responsible for turning encryption from an art to a science, and allowing better security in the internet age~\cite{goldwasser1984probabilistic}. Furthermore, she worked on a paper on zero-knowledge interactive proofs~\cite{goldwasser1989knowledge}, which makes it possible to prove a statement or concept without revealing any new information. These ideas were not directly applicable to answer the question posed by Rivest el al.~\cite{rivest1978data}, but it was a powerful tool for cryptographers to peek into the formalism of cryptography. Shafi Goldwasser was honored with the Turing Award, the highest honor in computer science, for her work in revolutionizing the field of cryptography. 

The story about how Goldwasser got interested in computer science is both inspiring and fascinating for all the women in STEM. She was born in 1959 in New York City. Her parents were Israeli, and her joint American/Israeli citizenship presaged the two countries that would play such an important role in her research. Her family returned to Israel, where Shafi attended grade school in Tel Aviv. In high school, she was especially interested in physics, mathematics, and literature. After her schooling, she returned to the U.S. and became an undergraduate in the mathematics department at Carnegie Mellon University (CMU). One computer science course that she especially remembers, taught by Jon Bentley, was an algorithms and discrete math course that she loved. Soon, however, she became interested in programming (which she had never done before) and computer science. She also worked on the CM* project at CMU, a 50-processor multiprocessor system. Shafi next had a summer internship at the RAND Corporation in Los Angeles that cemented her love for computer science.

Her interest in cryptography came a little late. After completing her undergraduate in 1979, she went on a vacation to California and decided to enroll in graduate school in computer science at the University of California, Berkeley, without knowing what she wanted to study. Her interaction with some of the computer scientists (Silvio Micali, Eric Bach, Manuel Blum, and others) during her graduate years in Berkeley initiated some interest in theoretical computer science, especially cryptography.  

The first problem Shafi began working on with Micali was how to hide partial information in ``mental poker''~\cite{goldwasser1984probabilistic}. Their solution was an essentially perfect way of encrypting a single bit (against a computationally limited adversary), and they invented a ``hybrid'' technique to show that independently encrypting individual bits causes the whole message to be secure. These proof were precursors to the proofs required for the theory of homomorphic encryption. 


After finishing up her Ph.D. from Berkeley, Shafi joined M.I.T. It was an exciting time when she came to M.I.T, as Ron Rivest, Micali, and Mike Sipser were still there. With Goldreich and Micali ~\cite{goldreich1986construct}, Shafi investigated whether the notion of a pseudorandom number generator could be generalized so that one could generate exponentially many bits (or equivalently, a function) pseudorandomly. This definition was in itself important, and is why we understand today what it means for a block cipher such as AES to be secure. They also showed how to provably transform a pseudorandom number generator into a pseudorandom function generator. These ideas had applications to the (then) new field of Learning Theory, providing examples of things that cannot be learned. These ideas were equally applicable to homomorphic encryption to show that the ciphertext didn't reveal any information about the raw data.

After working for a few years on ``Probabilistic encryption'' and ``pseudorandom functions'', her interest shifted towards Number theory~\cite{goldwasser1986almost}. It took decades for her to return to homomorphic learning. Just after Shafi switched her interests around 1995, Josh Daniel Cohen Benaloh came to the scene. In his Ph.D. dissertation ``Verifiable secret-ballot elections''~\cite{benaloh1989verifiable}, he developed theoretical tools and proofs for ensuring that the privacy of the voters remains intact though all communications are made public. He described a practical scheme for conducting secret-ballot elections in which the outcome of an election can be verified by all participants and even by non-participating observers. The theoretical tools developed for the secret-ballot election were equally applicable to allow a \textit{prover} to convince \textit{verifiers} that either statement \textbf{A} or statement \textbf{B} is true without revealing substantial information. His method of secret sharing homomorphisms enabled the computation of shared (secret) data and presented with a method of distribution of shares of secret such that each shareholder can verify the validity of all shares. 

Again, there were a few years of silence after the publication of Josh Benaloh's Ph.D. dissertation before the publication of ``Non-interactive crypto computing''~\cite{sander1999non} paper by Tomos Sander. 

Sander's paper focused on this classical problem:
\paragraph{Alice has an input x and Bob has function c. Alice should learn the value of C(x) but nothing else substantial about C. Bob should learn nothing else substantial about Alice's input x. } 

This problem can be viewed as computing over encrypted data where the parties do not engage in additional communication. Tomas Sander et al.~\cite{sander1999non} introduced a new protocol for secure evaluation of functions in polynomial time. The protocol involved one party sending encrypted input to a second party (crypto computer), which evaluates the input securely, and provides the output to the input party without learning the underlying function within a single round. This new protocol allowed the computation of logical AND operation using the encrypted data. Moreover, this scheme was homomorphic over a semigroup (instead of a group), thus expanding the range of algebraic structures that can be encrypted homomorphically. This was a huge accomplishment. Their protocol allowed some limited number of operations on the encrypted data. 

Just one year after the publication of ``Non-interactive crypto computing''~\cite{sander1999non}, Donald Beaver extended the protocol developed by Tomas Sander's team to include non-deterministic log-space functions~\cite{beaver2000minimal}. Moreover, Beaver introduced novel parallelization techniques to reduce the sub-computations on the functions to constant time complexity. This made the computation on the encrypted data faster, with reduced latency.

It took almost five years for another promising result. Dan Boneh, a Professor of the Computer Science Department at Stanford University published a landmark paper titled ``Evaluating 2-DNF formulas on ciphertexts''. In this paper, they presented a homomorphic public-key encryption scheme, that allowed the public evaluation of a 2-DNF formula ($\Phi$) on boolean variables ($x_1, x_2,\dots, x_n$), given the encryption (cyphertexts) of those boolean variables. 

In other words, their formulation allowed the evaluation of quadratic multivariate polynomials on ciphertext, provided the resulting value was within a small set. This formulation had the following major benefits: 1) it reduced the total number of communications required, 2) it facilitated the construction of a viable election system without the need of zero-knowledge proofs, and 3) it provided a protocol for universally verifiable computation. This was a keystone towards a plausible implementation of homomorphic learning. Dan Boneh then recruited a graduate student named Craig Gentry to extend their homomorphic scheme.

\begin{figure}
\centerline{\includegraphics[width=18.5pc]{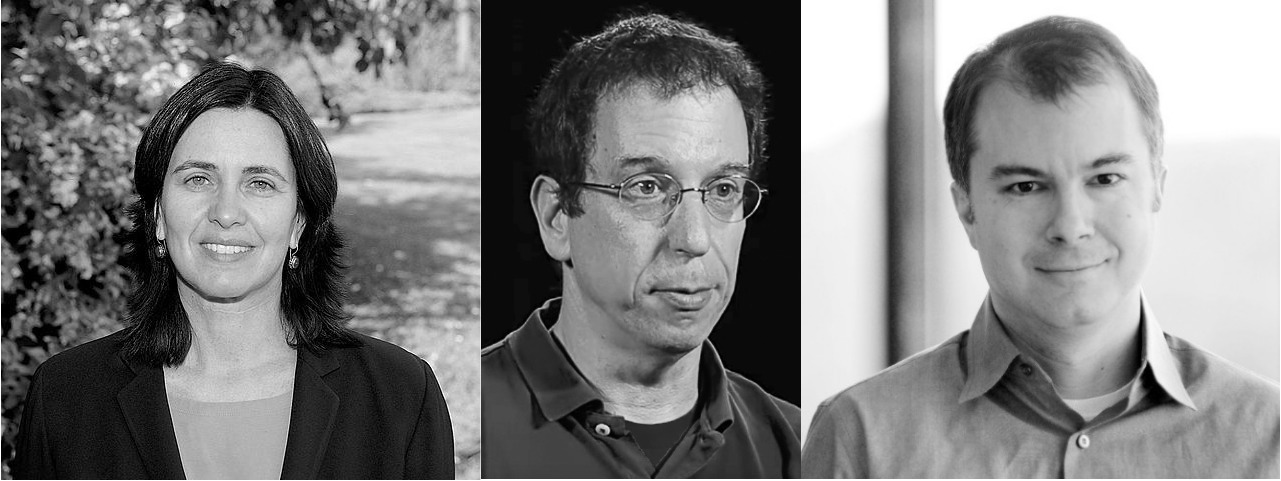}}
\caption{Fully homomorphic encryption pioneers. From left to right: Shafi Goldwasser, Dan Boneh, and Craig Gentry}
\label{fig:RivestQuestion}
\end{figure}

\section{The Promised Prince}

Craig Gentry appeared as the promised prince to fulfill a prophecy of making the dream of a fully homomorphic encryption (FHE) a reality. He worked with the Gödel Prize-winning Stanford Professor Dan Boneh and Turing award winner Shafi Goldwasser to come up with the first viable FHE scheme. 
In 2009, he published a plausible candidate construction of a fully homomorphic encryption~\cite{gentry2009fully} scheme, answering the problem posed in 1978 by Rivest and thought by many to be impossible to resolve. 

FHE makes it possible to perform arbitrary computations (mathematical operations like sum or product as well as more complicated operations) on encrypted data and without needing a secret key, and while keeping the individual pieces of data hidden and secure~\cite{gentry2009fully}.  For example, a web application could prepare an individual’s tax return using her encrypted financial information, without seeing any data in the raw form. Although FHE is not yet efficient enough for widespread implementation, Gentry's breakthrough has initiated a surge of new research globally in computer science as it has enormous implications for making the cloud computing environment more secure and compatible with data privacy for individuals. 


The path followed by Gentry to develop an FHE scheme started with viewing RSA encryption as  multiplicatively homomorphic: i.e, given an RSA public key and ciphertexts, one can efficiently compute a temporary ciphertext that encrypts the product of the original plain-texts. So, RSA is homomorphic concerning multiplicative operators. The encryption scheme that is homomorphic with any operator would enable arbitrary computation on encrypted data without the decryption key.   


The fully homomorphic encryption scheme~\cite{gentry2009fully} developed by Gentry is tied to a somewhat homomorphic scheme. His doctorate work shows that if the somewhat homomorphic encryption scheme is ``bootstrappable'', then a fully homomorphic encryption scheme can be built from it. Suppose there is an encryption scheme with a ``noise parameter'' (n) attached to each ciphertext. The noise parameter can be viewed as the random variable used in the encryption scheme, which generates different ciphertexts for the same raw text. The decryption works as long as the noise is less than some threshold $N >> n$. Furthermore, imagine if we have algorithms $Add$ and $Mult$ that can take ciphertexts $E(a)$ and $E(b)$ and compute $E(a+b)$ and $E(a*b)$, but at the cost of adding or multiplying the noise parameters. This immediately gives a ``somewhat homomorphic'' encryption scheme that can handle functions of depth roughly $(log \, log N - log \, log \,n)$. 

Now suppose that we have an algorithm \textbf{Recrypt} that takes a ciphertext $E(a)$ with noise $N' < N$ and outputs a fresh ciphertext $E(a)$ that also encrypts $a$, but which has noise parameter smaller than $\sqrt N$. This \textbf{Recrypt} algorithm is enough to construct a fully homomorphic scheme out of the somewhat homomorphic one. In his paper, the somewhat homomorphic encryption scheme is modified so that its decryption function has multiplicative depth at most $(log \, log N - log \, log \,  n - 1)$ i.e., less depth than what the scheme can handle. The somewhat homomorphic encryption scheme that has this self-referential property of being able to handle functions that are deeper than its own decryption function is known as ``bootstrappable'', and thus the somewhat homomorphic scheme which is ``bootstrappable'' is fully homomorphic.  

To get a feel for the actual working of the homomorphic scheme, let's go into details about a somewhat homomorphic encryption scheme on a toy problem. Let's look at a secret key encryption scheme which uses integers. The key is an odd integer $p > 2N$. An encryption of a bit $b$ is simply a random multiple of $p$, plus a random integer $B$ with the same parity as $b$, i.e $B$ is even if $b=0$ and is odd if $b=1$. More concretely, the ciphertext is \[c=b+2x+kp,\] where $x$ is a random integer in range ($-n/2, n/2$), and $k$ is an integer chosen from some range. Then we can decrypt the ciphertext by setting \[b \leftarrow (c \, mod \, p) \, mod \, 2,\] where $(c \, mod \, p)$ is the ``noise parameter'' in this scheme, which will be in the range $[-n, n]$. 
Decryption would work correctly as long as $(b+2x) \in [-N, N]$.


Let's take a concrete example and show that the simple homomorphic encryption scheme works. Suppose we want to encrypt the user's current balance on their account and find the total amount for all the users using the homomorphic encryption scheme for integers. Let the user's balance be represented by $b_1, b_2, \dots, b_n$. Specifically, let \[b_1=3, b_2=4, b_3=5\] and its corresponding binary representations are  \[b_1=00011, b_2=00100, b_3=00101\] respectively. Let the private key be $p=5$. Then $[-N, N] \subset (-p/2, p/2)$. Let $k$ be 2, which is inside the range $(-p/2, p/2)$. Let $x$ be a random integer inside $[-1,1]$. 
    
\begin{figure}
\centerline{\includegraphics[width=18.5pc]{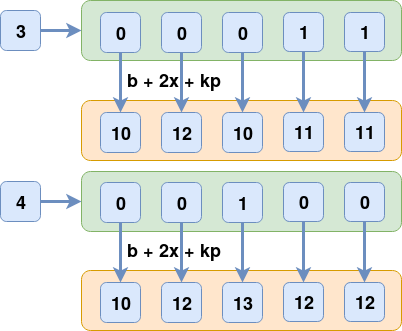}}
\caption{The encryption process for a simple homomorphic encryption scheme.}
\label{fig:encryption}
\end{figure}

\begin{figure}
\centerline{\includegraphics[width=18.5pc]{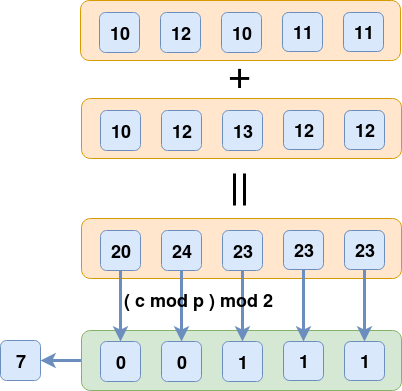}}
\caption{The decryption process for a simple homomorphic encryption scheme after the cyphertext obtained from encryption has been added.}
\label{fig:decryption}
\end{figure}

Figure~\ref{fig:encryption} describes how the cyphertext is generated for the integers. The integers are converted to binary and then encrypted using the cyphertext generating equation. The resulting ciphertexts can be operated on using any operator. For our example, we add the cyphertexts. The resulting cyphertext from the addition can be decrypted. Figure~\ref{fig:decryption} shows the process of decryption of the augmented cyphertext. The result of decrypting the augmented cyphertext is the same as the addition of the raw text. Thus, the simple homomorphic encryption scheme preserved the addition operation when applied to the cyphertext.

This toy example shows that the simple encryption is homomorphic when $N>>n$ and valid for $+,*$ operators. For these encryption schemes to be useful in the real world, we need fully homomorphic encryption that can operate on any arbitrary operators in a reasonable amount of time. Gentry's FHE supports arbitrary operators, at least in theory.
    
Gentry used the lattice-based approach in his doctorate dissertation to describe the first plausible construction for a fully homomorphic encryption scheme. His original cryptosystem reported the timing of about 30 minutes per basic bit operation, which is way too slow for real-world applications. Over the years, with extensive design and implementation work, the results have improved by many orders of magnitude. 

In 2010, Marten van Dijk, Craig Gentry, Shai Halevi, and Vinod Viakuntanathan~\cite{van2010fully} presented a second fully homomorphic encryption scheme based on the Gentry's construction that doesn't require ideal lattices. They showed that the ideal lattices could be replaced by a very simple somewhat homomorphic scheme that uses integers. This scheme was conceptually simpler than Gentry's ideal lattice scheme, but had similar properties with regards to homomorphic operations and efficiency.     

Gentry went on to tackle another longstanding open problem, moving beyond cryptographic bilinear maps (which have been used widely in encryption, cryptographic digital signatures, and key agreement) to multilinear maps. 
He and his collaborators were the first to publish a plausible candidate multilinear map. In 2013, he and his colleagues leveraged those findings to present the first example of cryptographic software obfuscation, a discovery that is paving the way for the encryption of entire programs while keeping their functionality intact, thereby making it infeasible to reverse engineer the program~\cite{garg2014fully}. This is one solid practical example of homomorphic encryption. He is inspiring not only a flood of practical applications but also opening doors for new intellectual pursuits across the whole of cryptography.


\section{The Opensource Factory}

The publication of Gentry's doctorate dissertation in 2009 made the fully homomorphic encryption (FHE) a buzz word. Still, FHE was not computationally efficient to be applied to real-world applications. Over the last few years, with extensive research and opensource tools like \textit{Seal}~\cite{laine2016simple}, \textit{Palisade}~\cite{hallman2018building}, \textit{HElib}~\cite{halevi2014algorithms} and others started to change the landscape of the homomorphic encryption. The integration of these tools into popular Machine Learning libraries such as TensorFlow and PyTorch enabled amateurs and researches alike to build applications focused on privacy. These opensource tools are responsible for the homomorphic encryption to reach the masses. Still, only a few specialized companies provide homomorphic learning as a service to healthcare and government organizations, and it might take a few more years for homomorphic learning to be ubiquitous among the cloud service providers.    

Secure multi-party computation, differential privacy, and federated learning are the popular form of privacy-focused learning. In 2018/2019, there were more than ten privacy-focused learning library released as open-source software. Secure multi-party computation is all about computing a function jointly by different parties over their inputs while keeping those inputs private. Shafi, with Yael Tauman Kalai and Guy Rothblum, introduced ~\cite{goldwasser2015delegating} one practical formulation of secure multi-party computation, and showed how to efficiently delegate the computation of small-depth functions to the untrusted ``cloud''.
    
Differential privacy addresses the paradox of learning nothing about an individual while learning useful information about a population. In general, differential privacy is not an algorithm but a definition that outlines ways to perform privacy-preserving data analysis ~\cite{dwork2014algorithmic}. In a simplified differential privacy model of computation, we assume the existence of a trusted and trustworthy curator who holds the data of individuals in a database \textbf{D}, typically comprised of some number \textit{n} of rows, and each row contains the data of a single individual. The differential privacy goal employs a \textit{mechanism} to simultaneously protect every individual row while permitting statistical analysis of the database as a whole. This \textit{mechanism} is an algorithm that takes as input a database, the set of all possible database rows, random bits, and a set of queries, and produces an output string. The output string can be decoded to produce  relatively accurate answers to the queries without revealing the information of the individuals. Tang et al.~\cite{tang2018homomorphic} combined homomorphic encryption with differential privacy to create a novel privacy-preserving algorithm \textit{Heda} to train an ML classifier on the UCI ML datasets~\cite{Dua2019}.

Federated learning is a distributed architecture for learning a predictive model using decentralized mobile devices or servers, keeping the local data samples private, and only publishing the learned updates to the central server using homomorphic encryption. 
Federated learning allows for personalization, scaled deployment of models, lower latency, and less power consumption, all while ensuring data privacy and security. One popular success example of federated learning is the ``Gboard on Android'' (the Google Keyboard). When the Gboard shows a suggested query, it is using federated learning on the history on-device and updating it based on whether you clicked the suggestion or not~\cite{brendan2017}. 


Currently, there are a few frameworks like PySyft~\cite{ryffel2018generic} and HE Transformer~\cite{boemer2019ngraph} that allow the users to perform homomorphic learning with Neural Network models on encrypted data. These frameworks decouple private data from model training using Federated Learning, Differential Privacy, and Multi-Party computation (MPC) within the main deep learning frameworks like PyTorch and TensorFlow.

One of the influential people who is taking the responsibility of developing opensource tools for integrating fully homomorphic encryption (FHE) and ML libraries is Andrew Trask. He is a researcher pursuing a Doctorate at Oxford University, where he focuses on Deep Learning with an emphasis on human language. Dowlin et al.~\cite{gilad2016cryptonets} were the first to train a neural network model on encrypted MNIST data with high throughput and accuracy using the SEAL~\cite{laine2016simple} library for homomorphic encryption. Hesamifard et al.~\cite{hesamifard2017cryptodl} trained a deep neural network over encrypted MNIST and CIFAR data with high efficiency and accuracy. Badawi~\cite{badawi2018alexnet} et al. trained the first homomorphic convolution neural network (HCNN) on encrypted data with Graphics Processing Unit (GPU)s. All previous approaches before HCNN ran only on the CPU. HCNN is truly remarkable, as it achieved the high-security level and great classification accuracy for MNIST and CIFAR.

For the machine learning enthusiasts and practitioners, homomorphic learning provides an exciting opportunity to democratize AI, and a privacy-focused way to solve large-scale ML problems. Homomorphic learning is a nascent field with lots of hope and promises, eagerly waiting for next Craig Gentry to revolutionize the field.




\section{CONCLUSION}
This article described a brief history on the development of homomorphic encryption schemes, highlighting the collective contributions of a cohort of brilliant minds. It took more than thirty years of communal effort to find a theoretical answer to the question ``Can we compute upon encrypted data?'' posed by Rivest et al~\cite{rivest1978data}. It took another ten years to develop efficient homomorphic encryption frameworks that could be coupled with other machine learning libraries. Multi-party computation and federated learning aided with homomorphic encryption have already started making a powerful impact on privacy-focused machine learning communities. It might still take a few years before homomorphic learning is ubiquitous.

\section{ACKNOWLEDGMENT}
I would like to thank Dr. Sean Warnick for giving us this opportunity to work on this project for CS611 course. Also, I would like to thank Connor Anderson, Michael DeBuse, and Aatish Neupane for their constructive feedback and suggestions to make this paper better.

\bibliographystyle{ieeetr}
\bibliography{biblography}


\end{document}